# Theory of Spin Transport Across Domain-Walls in (Ga,Mn)As

Rafał Oszwałdowski*, Jacek A. Majewski¶, and Tomasz Dietl¶§

*Institute of Physics, Nicolaus Copernicus University, Grudziądzka 5, PL 87-100 Toruń, Poland*
*¶ Institute of Theoretical Physics, Warsaw University, PL 00-681Warszawa, Poland*
*§ Institute of Physics, Polish Academy of Sciences and ERATO Semiconductor Spintronics Project of Japan Science and Technology Agency, PL 02-668 Warszawa, Poland*

**Abstract.** We present results of numerical calculations of domain-wall resistance in the ferromagnetic semiconductor (Ga,Mn)As. We employ Landauer-Büttiker formalism and the tight binding method. Taking into account the full valence band structure we predict the magnitude of the domain-wall resistance without disorder and compare it to experimental values. Next we add disorder to the model and study numerically both small and large disorder regime.

**Keywords:** dilute magnetic semiconductors, magnetic domains, disorder
**PACS:** 75.47.Jn, 72.25.Dc, 75.50.Pp

## INTRODUCTION

Recently, current-induced domain-wall displacement [1] and intrinsic domain-wall resistance [2,3] have been observed in ferromagnetic semiconductor (Ga,Mn)As. These results attract a lot of attention as they address the question of spin dynamics in the presence of spatially inhomogeneous spin texture as well as they open the doors for novel concepts of high density memories and logic devices.

We study intrinsic domain-wall resistance (DWR) in perpendicular anisotropy (Ga,Mn)As films employing recent theory that combines an empirical multi-orbital tight-binding approach with a Landauer-Büttiker formalism for spin transport [4]. In this scheme, the entire complexity of the valence band structure is taken into account. Since the mean free path of the carriers is typically shorter than the domain wall (DW) width, carrier scattering has to be taken into account. This is accomplished within the 1D Anderson model by assuming random values of on-site energies.

## METHOD AND RESULTS

We assume the in-plane current to flow along the x-axis, and the Bloch DW profile to be described by magnetization $M(x)$ according to: $M_x=0$; $M_y=1/\cosh(x/\lambda_W)$; $M_z=\tanh(x/\lambda_W)$, where $\lambda_W$ denotes the DW length parameter.

## Band Structure Contribution to the DWR

It was found in early studies [5] that the resistance of DWs vanishes in the adiabatic limit $\lambda_W k_F \gg 1$. However, in the presence of spin-orbit coupling, DWR is non-zero even in the adiabatic limit [6]. This asymptotic DWR must be taken into account in (Ga,Mn)As, where the DW width is of the same order as Fermi wavelength [7]. In Fig. 1 we present our results for (Ga,Mn)As DWR for relevant values of Mn content and Fermi level $E_F$. Calculations have been performed for $\lambda_W = 5$ nm, which corresponds to the predicted DW width [8].

However, even the highest value of Fig. 1 is an order of magnitude lower than the DWR determined experimentally [3]. Thus, the measured value of DWR cannot be explained by assuming ballistic transport, where rotating magnetization is considered as the only source of scattering.

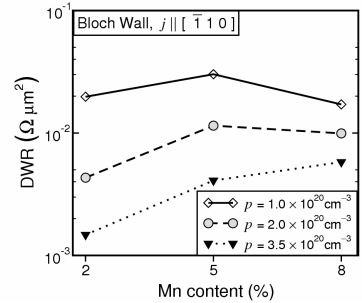

**FIGURE 1.** Domain-wall resistance in (Ga,Mn)As calculated for valence band structure of GaAs but neglecting disorder.

## Disorder Effects in DWR

Given the short mean free path $l$ in (Ga,Mn)As, comparable to $\lambda_W$ and $k_F^{-1}$ [9], a large role of disorder can be anticipated. Apart from practical implications, this issue is interesting theoretically, since negative DWR in the presence of disorder was predicted theoretically [10,11]. We neglect band structure effects assuming a simple parabolic band, but now introduce random on-site energies in particular atomic layers according to the time-honored 1D Anderson model [12]. The disorder is characterized by the width $d$ of the random energy distribution. For a given disorder configuration, adopting experimentally relevant values of the ratio of spin splitting $\Delta$ to Fermi energy $E_F$, we calculate DWR as a difference between resistance with and without DW. Then we perform averaging for many disorder configurations, and use root-mean-square value as a measure of DWR variance. Using the Drude formula we obtain from the calculated resistivity (without DW) the value of $k_F l$, which corresponds to a particular magnitude of $d$.

We find, as could be expected, that disorder has no effect on DWR as long as $l \gg \lambda_W$. However, when $d$ increases and thus $l$ decreases, an additional effect, is observed. It is well known, [12] that the variance of resistance calculated in the Anderson model, steeply grows for increasing disorder.

In Fig. 2 we show that the rms of the DWR quickly increases. Each point was calculated averaging over 100 disorder configurations. For values of $d$ yielding $k_F l < 100$, the relative DWR error is larger than one for any spin-splitting $\Delta$. Thus, in this experimentally relevant regime, reliable predictions based on configurational averaging over 1D disorder would require a very high number of configurations.

It is interesting to use the same approach to investigate the possibility of negative DWR in the large disorder limit. To compare more directly with the work of Jonkers *et al.*, [11] we use $k_F \lambda_W / \pi \approx 4$, and $\Delta/E_F \approx 0.8$ in our calculations. We denote the transmission coefficient for the structure with (without) a DW by $T$ ($T_0$). We calculate $T$ and $T_0$ for $k_{||} = 0$.

In agreement with the results of [11] we find that for some disorder configurations $T/T_0 > 1$. Percentage of such configurations grows with increasing disorder. This suggests that under some conditions it is possible to obtain, on average, $T/T_0 > 1$ for majority of the tunneling channels $k_{||}$.

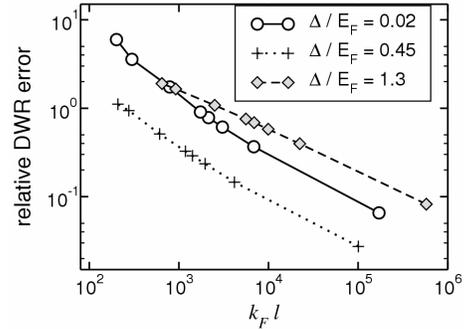

**FIGURE 2.** Dependence of relative variance of domain-wall resistance on the mean free path $l$. The DW width is of the order of 10 nm. For the shortest $l$ (highest disorder) also the r-m-s error fluctuates. Note the log-log scale.

If this were the case, a negative DWR would be realized. We must note, however, that with increasing $d$, the variance of $T/T_0$ increases, yielding the calculated average uncertain.

## ACKNOWLEDGMENTS

This work was partly supported by EU NANOSPIN project (EC:FP6-2002-IST-015728). R.O. acknowledges an additional support by Nicolaus Copernicus University Grant 381-F.